\pdfoutput=1
\RequirePackage{fix-cm}
\documentclass[smallextended]{svjour3}     
\smartqed 
\usepackage{graphicx}
\usepackage{mathptmx}     
\usepackage[numbers]{natbib}
\usepackage{amsfonts}

\newcommand{\ket}[1]{\left| #1 \right>} % for Dirac bras
\newcommand{\bra}[1]{\left< #1 \right|} % for Dirac kets
\newcommand{\braket}[2]{\left< #1 \vphantom{#2} \right| \left. #2 \vphantom{#1} \right>} % for Dirac brackets

 \journalname{Quantum Information Processing}

\begin{document}

\title{The quest for a Quantum Neural Network%\thanks{Grants or other notes
}

\author{Maria Schuld \and  Ilya Sinayskiy \and Francesco Petruccione}

\institute{M. Schuld \at
             Quantum Research Group, School of Chemistry and Physics,  University of KwaZulu-Natal, Durban, KwaZulu-Natal, 4001, South Africa
             \email{schuld@ukzn.ac.za}           
           \and
           I. Sinayskiy \at
               Quantum Research Group, School of Chemistry and Physics,  University of KwaZulu-Natal, Durban, KwaZulu-Natal, 4001, South Africa and\\ National Institute for Theoretical Physics (NITheP), KwaZulu-Natal, 4001, South Africa 
\and
 F. Petruccione \at
               Quantum Research Group, School of Chemistry and Physics,  University of KwaZulu-Natal, Durban, KwaZulu-Natal, 4001, South Africa and\\ National Institute for Theoretical Physics (NITheP), KwaZulu-Natal, 4001, South Africa 
}

\date{Received: date / Accepted: date}

\maketitle

\begin{abstract}
With the overwhelming success in the field of quantum information in the last decades, the `quest' for a Quantum Neural Network (QNN) model began in order to combine quantum computing with the striking properties of neural computing. This article presents a systematic approach to QNN research, which so far consists of a conglomeration of  ideas and proposals. It outlines the challenge of combining the nonlinear, dissipative dynamics of neural computing and the linear, unitary dynamics of quantum computing. It establishes requirements for a meaningful QNN and reviews existing literature against these requirements. It is found that none of the proposals for a potential QNN model fully exploits both the advantages of quantum physics and computing in neural networks. An outlook on possible ways forward is given, emphasizing the idea of Open Quantum Neural Networks based on dissipative quantum computing. 

\end{abstract}

\section{Introduction}\label{intro}

Quantum Neural Networks (QNNs) are models, systems or devices that combine features of quantum theory with the properties of neural networks. Neural networks (NNs) are models of interconnected units based on biological neurons feeding signals into one another. A large class of NNs uses binary McCulloch-Pitts neurons  \cite{rabinovich06,mcculloch43}, thus reducing the complex process of signal transmission in neural cells to the two states `active/resting'. The analogy with the two-level qubit serving as the basic unit in quantum computing gives an immediate connection between NN models and quantum theory. The majority of proposals for QNN models are consequently based on the idea of a qubit neuron (or `quron' as we suggest to name it), and theoretically construct neurons as two-level quantum systems.\\

Although close to discussions about the potential `quantumness of the brain' \cite{kak95,freeman08,hameroff98}, QNNs do not intend to explain our brain functions in terms of quantum mechanics. Neurons are macroscopic objects with dynamics on the timescale of microseconds, and a quron's theoretically introduced two quantum states refer to a process involving millions of ions in a confined space, leading to estimated decoherence times in the order of  $10^{-13}\; \mathrm{sec}$ and less \cite{tegmark00}, thus making quantum effects unlikely to play a role in neural information processing. However, QNNs promise to be very powerful computing devices \cite{panella11,rigatos07}. Their potential lies in the fact that they exploit the advantages of superposition-based quantum computing and parallel-processed neural computing at the same time. QNN research can furthermore be seen as a part of a growing interest of scientist and IT companies to develop quantum machine learning algorithms for efficient big data processing \cite{lloyd13,aimeur13,wiebe14}. Artificial neural networks thereby play an important role as intelligent computational methods for pattern recognition and learning.\\

The debate on quantum approaches to neural networks emerged in the wake of a booming research field of quantum computing two decades ago. One year after Shor proved the potential power of quantum computers by introducing his famous prime factorisation algorithm in 1994 \cite{shor97}, Kak \cite{kak95} published some first ideas to find a junction between neural networks and quantum mechanics. Since then, a number of proposals claim the term `Quantum Neural Network' \cite{andrecut02,altaisky01,gupta01,behrman02,fei03,zhou12,oliveira08,toth00}. However, until today a major breakthrough is still outstanding and QNN research remains an exotic conglomeration of different ideas under the umbrella of quantum information. The reason for this is that beyond the `quron', the nonlinear dissipative dynamics of neural computation \cite{rabinovich06} are fundamentally different to the linear, unitary dynamics of quantum computing \cite{nielsen10}. To find a meaningful QNN that integrates both fields into the `quantum evolution of a neural network' is therefore a highly nontrivial task.\\

This article presents a systematic study of QNN research.\footnote{The only systematic review in the field of QNNs was given in 2000 by Ezhov and Ventura \cite{ezhov00} (not counting the brief overview of the different types of implementations of QNNs in Oliveira et al. \cite{oliveira08}). To our knowledge, there is no recent comprehensive review.} First, neural computing as well as quantum computing are briefly introduced to sketch the above mentioned problem structure (Sections \ref{nc} and \ref{qco}). Second, a framework of requirements for a potential QNN based on Hopfield-like neural networks with the property of associative memory is established and existing literature is reviewed on this background (Section \ref{rev}). It is found here that none of the proposals for a QNN model satisfies the given requirements. In conclusion, new ideas for approaches to QNN models are needed. This article therefore gives an outlook on the idea of Open Quantum Neural Networks (Section \ref{dis}). Based on the theory of open quantum systems \cite{breuer02} and the new emerging field of dissipative quantum computing \cite{verstraete09}, Open Quantum Neural Network models would make use of dissipation in order to obtain dynamical properties similar to neural networks.\\

\section{Neural computing}\label{nc} 

\begin{figure}
                \includegraphics[width=\textwidth]{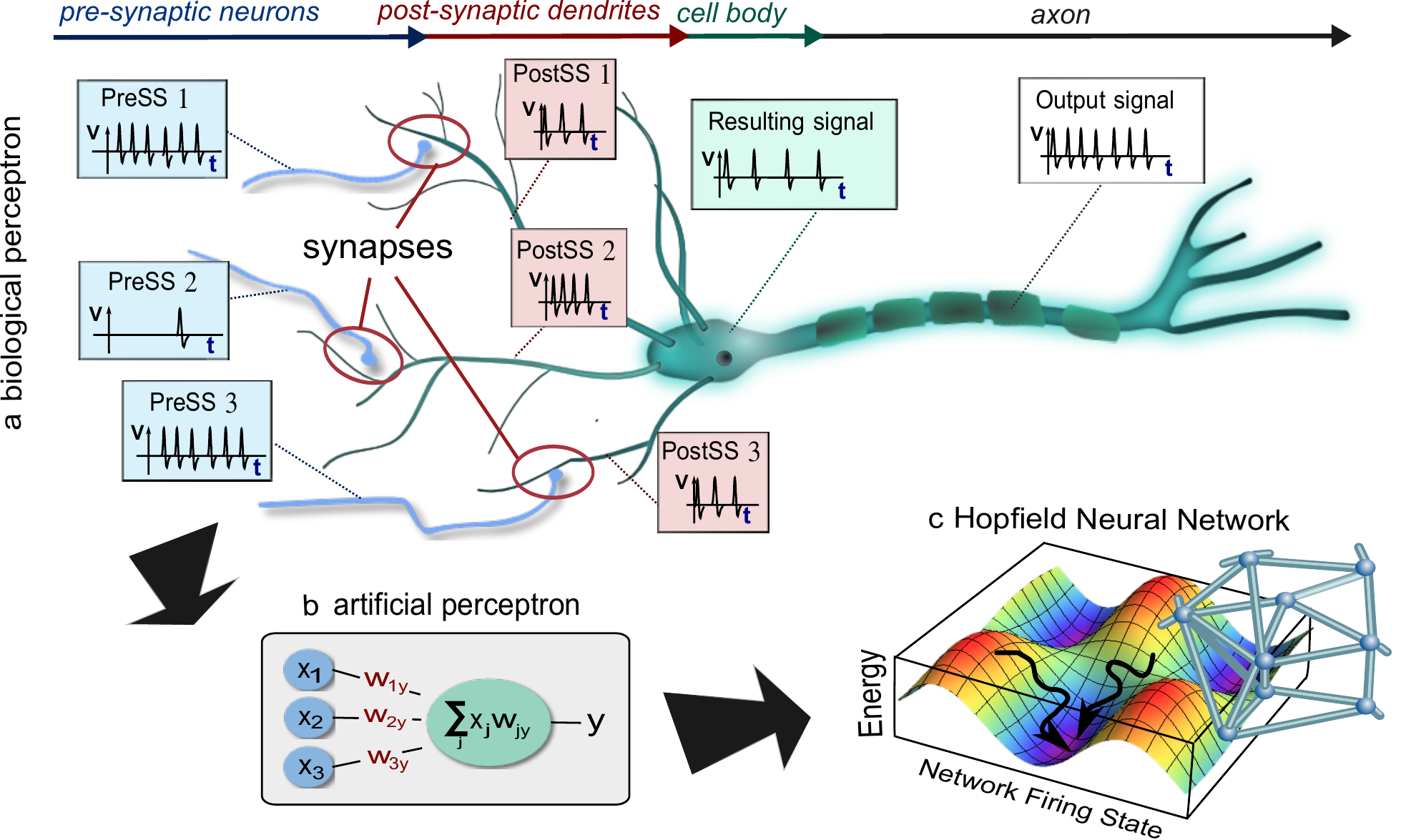}
              \caption{Illustration of the perceptron mechanism in neural computing.(a) In this illustration of a biological perceptron, three `input neurons' (blue) feed their pre-synaptic signals into the `output neuron' (green). The pre-synaptic signals are transformed into post-synaptic signals by synapses (red) and travel through the dendrites to the cell body where they add up to a resulting signal. The output signal that propagates along the axon and feeds into the next layer of neural cells nonlinearly depends on the strength of the resulting signal. (b) Artificial perceptrons are based on binary neurons of the states active/resting, represented by the values $-1,1$. The input neurons are denoted by $x_1,x_2, x_3 \in \{-1,1\}$, the synaptic weights are chosen to be  $w_{1y}, w_{2y}, w_{3y} \in [-1,1]$ and $y$'s output is $-1$ or $1$,  depending on the resulting signal $\sum_j x_j w_{jy}$. (c) This simplified perceptron mechanism leads to rich dynamics in fully connected neural networks (here illustrated by a graph), where the global firing state of a neural network converges to local attractors, a feature that gives rise to associative memory, pattern classification etc.}
  \label{NC}
\end{figure}

Computing in artificial neural networks is derived from our neuroscientific understanding of how the brain processes information in order to master its impressive tasks. The brain is widely believed to encode information in the connectivity architecture of $10^{11}$ neural cells connected by $10^{14}$ synapses \cite{purves08}. Neurons propagate firing signals along this architecture. These so called \textit{action potentials} are traveling depolarisations of the equilibrium membrane potential due to ion flux through voltage-gated channels along their axons. The signals feed into other neurons through synaptic connections between the axon ends of a pre-synaptic neuron and the dendrites of a post-synaptic neuron (see Fig. \ref{NC}). In the simplified model of a McCulloch-Pitts neuron \cite{mcculloch43}, a neuron is in an `active' state if it is firing with a sufficient rate, while otherwise it is `resting'. In mathematical notation, neurons are symbolised by variables $x,y = \{-1,1\} $, where `$1$' indicates that the neuron is firing and `$-1$' that it is resting. \\

The activation mechanism of a neuron $y$ due to the input of $N$ other neurons $x_1,...,x_m$ forms the core of neural computing. This setup is called a \textit{perceptron} (see Fig. \ref{NC}). Two important features characterise this mechanism. First, the incoming (`pre-synaptic') signal $\{-1,1\}$ from each neuron $x_1,...,x_m$ is transformed into a post-synaptic signal by the respective synapse connecting it to neuron $y$. This amplification of incoming signals ensures a rich computational variety \cite{abbott04} and the properties of a NN are indeed stored in the synapses. The synapse's modulating effects can be formally expressed through a parameter $w_{iy}\in [-1,1], \; i = 1,...,m$ symbolising the synaptic strength. Second, neuron $y$ is activated in an `integrate-and-fire' mechanism, meaning that the post-synaptic signals are simply added up and compared with the specific threshold $\theta_y$ of neuron $y$. If the resulting signal exceeds the threshold, $y$ is activated and consequently set into the state `fire'; if not, $y$ rests. This can be formulated through the equation

\begin{equation} y =  \left\{   \begin{array}{l l}
		    1, & \quad \mathrm{if} \; \sum\limits_{j=1}^m w_{jy} x_j \leq \theta_y,\\
   		    -1, & \quad \mathrm{else.} 
  		\end{array} \right . \label{afperc} \end{equation}

A neural network is then a set of neurons $\{x_1,...,x_N\} \in\{-1,1\}$ with respective thresholds $\{\theta_1,...,\theta_N\}$, connected through synaptic strengths $w_{ij}\in [-1,1], \; i,j = 1,...,N$. Note that a network can thus encode a binary string  $(x_1 = \pm1,...,x_N= \pm1)$ in its firing pattern or `network state'. Each neuron and its input from adjacent neurons form a perceptron unit. An update of neuron $x_i, \; i \in \{1,...,N\}$ follows the activation function and reads
\begin{equation} x_i =  \left\{   \begin{array}{l l}
		    1, & \quad \mathrm{if} \; \sum\limits_{j=1}^N w_{ji} x_j \leq \theta_i,\\
   		    -1, & \quad \mathrm{else.} 
  		\end{array} \right . \label{af} \end{equation}
 The neurons of a neural network can be updated in different protocols, for example in a synchronous, chronological or random order \cite{hopfield82,rojas96}.\\

A neural network of McCulloch-Pitts neurons in which the connectivity architecture obeys 
\begin{equation} w_{ij} = w_{ji}, \qquad w_{ii} = 0, \label{ws} \end{equation}
is called \textit{Hopfield Neural Network (HNN)} \cite{hopfield82,hopfield86}. We will concentrate on HNNs in the following, although much of what is said can be easily transferred to the other big class of NNs, so called \textit{Feed-Forward Neural Networks}. Although of a simple setup, the Hopfield model  shows the powerful feature of associative memory. Associative memory is the ability to retrieve the network state out of $P$ stored network states $X^P = \{(x_1^{(1)},...,x_N^{(1)}),...,(x_1^{(P)},...,x_N^{(P)})\}$ which is closest to the input pattern in terms of Hamming distance\footnote{The Hamming distance is the number of state flips to turn one binary string into another one, thus measuring the overlap between two binary strings \cite{hamming50}.}. Also called `content addressable memory', associative memory allows to compute incomplete input information instead of the pattern's exact storage address needed in a computer's Random Access Memory. Accessing stored information upon incomplete inputs forms the basis of human memory and learning. \\

The easiest way to understand how HNNs store and retrieve information in terms of network firing states is to introduce the energy function of a firing state $(x_1,...,x_N)$ with threshold vector $(\theta_1,...,\theta_N)$ and synaptic connections $w_{ij}, \;\;  i,j = 1,...,N$, which reads 

\[ E(x_1,...,x_N) = - \frac{1}{2} \sum\limits_{i=1}^{N} \sum\limits_{j=1}^{N} w_{ij} x_i x_j +\sum\limits_{i=1}^{N} \theta_{i} x_i . \]

Note that this is equivalent to the energy function of an Ising spin-glass model, a fact that opens HNN research to methods of statistical physics \cite{amit85,hemmen86}. Similar to a relaxing spin chain, the Hopfield network's dynamics converge to minimum or ground state of the energy function. In other words, each update of a neuron minimises the energy of the network state if possible, or maintains it otherwise \cite{rojas96}. Memorised firing patterns are thus stored as stable attractors in the NN's dynamics and an initial firing pattern will always end up in the minimum of the basin of attraction it lies in \cite{rabinovich06}.\\

HNNs inherit the attractors from the nonlinear activation function given in Eq. (\ref{af}). The property $w_{ii} = 0$ makes sure that all attractors are stable states (as opposed to limit cycles of alternating states) \cite{rojas96}. To store the set of firing patterns $X^P = \{(x_1^{(1)},...,x_N^{(1)}),...,(x_1^{(P)},...,x_N^{(P)})\}$ in the energy landscape of a HNN, the synaptic weights can be chosen according to Hebb's learning rule
\begin{equation} w_{ij} =  \frac{1}{P} \sum\limits_{p=1}^{P} x_i^{(p)} x_j^{(p)}, \label{hebb} \end{equation}  
reflecting that neurons that have the same state in the majority of memory patterns will receive a synaptic weight close to $1$ while a high antiparallel correlation gives rise to a weight close to $-1$. The number of patterns that can be stored in a network largely depends on the patterns themselves, for example patterns with a low Hamming distance are more prone to be confused in the retrieval process. This is why different upper bounds for the storage capacity can be found. Hertz et al. estimate a capacity (memorisable patterns divided by the number of neurons) of $c = 0.138$ \cite{hertz91}, while others speak of $c =N / (4 \ln N)$ \cite{mehrotra96}.\\

Hopfield published a variation of his model in 1984 \cite{hopfield84} based on so called graded-response neurons. Instead of the binary values of a McCulloch-Pitts neuron, graded response neurons can take values out of a continuous range, for example $x \in [-1,1]$. The step-function gets replaced by a sigmoid function
\[\mathrm{sgm}(a; \kappa) = \frac{1}{1+ \mathrm{e}^{-\kappa a}}, \]
with steepness parameter $\kappa$ and the updaing or activation function of a neuron $x_i$ due to the input of neurons $x_j, \; j \in \{1,...,N\}$ consequently reads
\begin{equation} x_i =  \mathrm{sgm}\left(\sum\limits_{j=1}^N w_{ji} x_j; \kappa \right). \label{gr} \end{equation}
Note that with $\kappa \rightarrow \infty$, the sigmoid function includes the step-function as a limit case. Hopfield showed that the graded-response model has equivalent attractor-properties to the original model.

\section{Quantum computing} \label{qco}

The term `quantum computing' usually refers to the engineered coherent evolution of a number of quantum two-level systems of a 2-dimensional Hilbert space $\mathcal{H}_2$ with basis $\{\ket{0}, \ket{1}\}$ called qubits and  described by the wavefunction
\[\ket{\psi} = \alpha \ket{0} + \beta \ket{1}, \]
\[|\alpha|^2 + |\beta|^2 = 1 , \;\; \alpha, \beta \in \mathbb{C}.\] 
The evolution of qubits through so called quantum channels is carried out by unitary quantum gates that manipulate the qubits just like classical bits are manipulated by logic gates in a computer \cite{nielsen10,divincenzo00}. In mathematical language, quantum gates are unitary linear transformations of the quantum state vectors. The power of quantum computing lies in the fact that a qubit -- as opposed to a classical bit -- can be in a superposition of its two basis states. This allows for the parallel exploitation of different `paths' of computation at the same time using interference effects. The result of a quantum computation algorithm is of probabilistic nature and encoded in the final coefficients $|\alpha|^2 $ and $|\beta|^2$ that indicate probabilities to measure either one of the two basis states. It can be read out through repeated measurements of the qubit system. \\

\begin{figure}
\centering
                \includegraphics[width=0.7\textwidth]{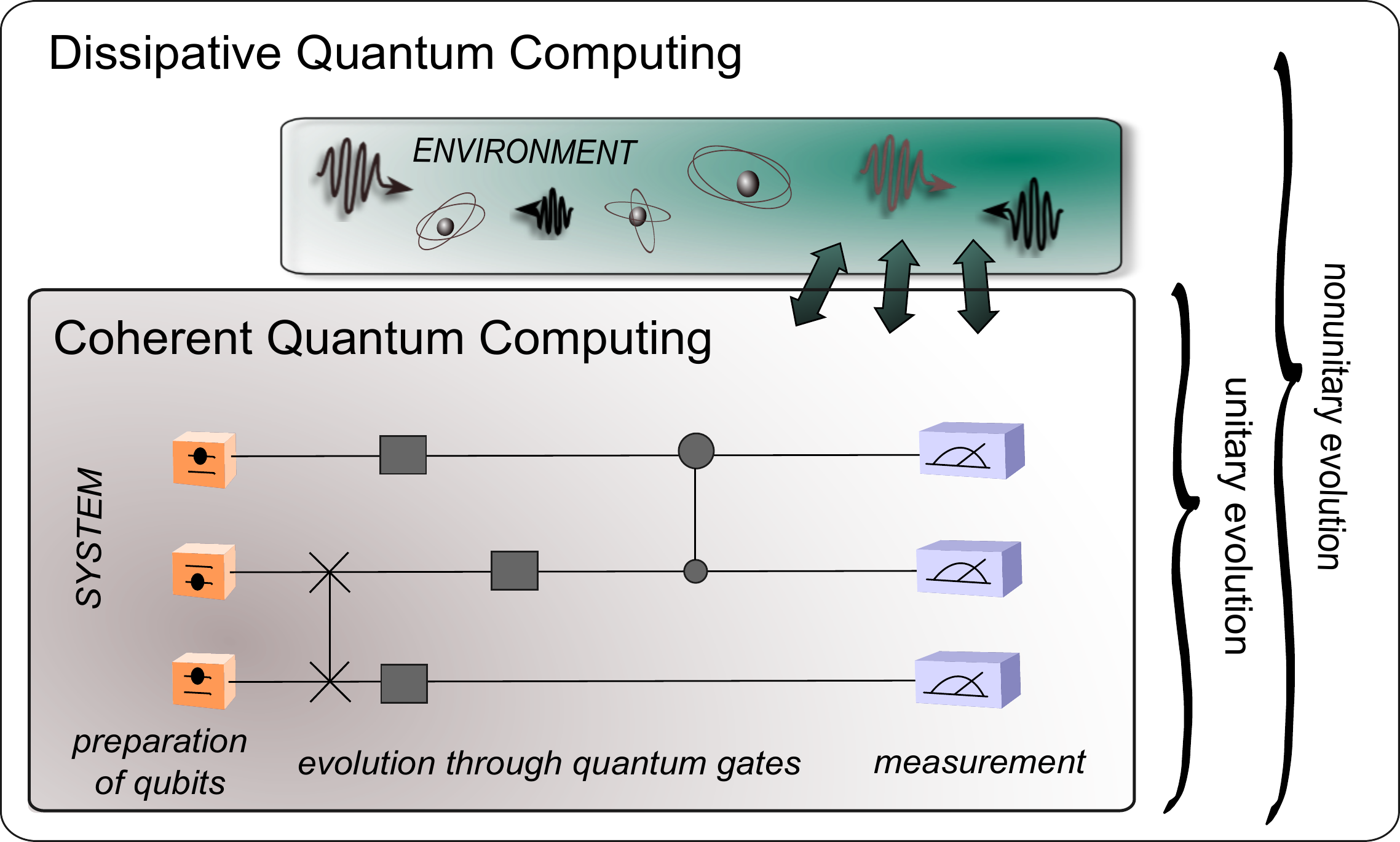}
              \caption{Quantum computing consists of the preparation, evolution and measurement of qubits. In regimes of dissipative quantum computing, the qubits are not only manipulated by unitary quantum gates, but also interact with an environment }
  \label{qc}
\end{figure}

Quantum physics is a linear theory, meaning that maps of one state onto another are executed by linear operators. The operators are furthermore unitary, ensuring probability conservation. In fact, the current challenge in the implementation of a quantum computer lies in the difficulty to maintain coherence in a multiple-qubit system in order to simulate unitary dynamics. Real systems interact with an environment, thus introducing effects of dissipation (loss in the population of quantum states) and decoherence (destruction of quantum state correlations). However, in recent years more general forms of quantum computing have been introduced that do not regard decoherence or dissipation as necessary evils, but as a means to engineer the desired evolution of a quantum system. So called \textit{Dissipative Quantum Computing} \cite{verstraete09} is based on the theory of open quantum systems describing systems in interaction with a large environment. The total system consisting of the principal system plus its environment still obeys the unitary evolution of quantum theory, but the principal system alone propagates nonunitarly and is exposed to decoherence. In the scheme found by Verstraete, Wolf and Cirac \cite{verstraete09}, the system and environment is engineered so that the initial state $\ket{0,...,0}$ is mapped to a desired final state $\ket{\psi_T}$ that can be read out of the system after time $T$ with a probability of $\frac{1}{T}$. \\

The idea of dissipative quantum computing is highly interesting for QNN research, since it allows for quantum computing algorithms based on dynamic attractors and steady states. It is of course obvious that the scheme above \cite{verstraete09} does not realise associative memory since the choice of inputs and output is limited to one each. However, other ideas of open quantum systems with more than one ground state and attractive dynamics could be a way forward in QNN research that shall be discussed further below. Other alternatives to coherent quantum computing that could be highly interesting for QNN research are measurement-based quantum computing \cite{briegel09},  adiabatic quantum computing \cite{farhi00} or duality quantum computing \cite{long11}.

\section*{Requirements for a QNN model} \label{req}%1p

As mentioned earlier, the basic idea of introducing quantum properties into classical NNs is to replace the McCulloch-Pitts neuron $x = \{-1,1\}$ by a `quron' $\ket{x}$  of the two-dimensional Hilbert space $\mathcal{H}^2$ with basis $ \{\ket{0}, \ket{1}\}$. The state $\ket{\psi}$ of a network with $N$ qurons thus becomes a multiparticle quantum state of the $2^N$-dimensional Hilbert space $\mathcal{H}^{2^N} =\mathcal{H}^2 \otimes \dots \otimes  \mathcal{H}^2$ with basis $\{\ket{0,0, \dots ,0}, ..., \ket{1,1, \dots ,1}  \}$  reading
\begin{equation}\ket{\psi} =  \sum \limits_{i = 1}^{2^N} a_i \ket{x_1 x_2 \dots x_N}_i, \label{nws} \end{equation}
where the $a_i$, $i\in \{1,...,2^N\}$ refer to the complex amplitudes assigned to the respective network basis states.\\

\begin{figure}
\centering
                \includegraphics[width=0.4\textwidth]{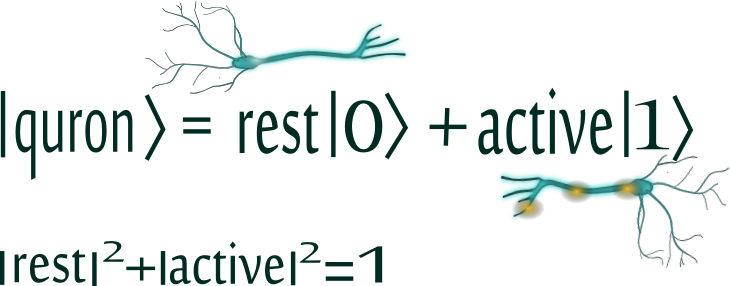}
              \caption{A quron is a qubit in which the two levels stand for the active and resting neural firing state. This allows for a neural network to be in a superposition of firing patterns, the central property that is exploited in QNN models }
  \label{quron}
\end{figure}

Apart from the quron, proposals for QNN models vary strongly in their proximity to the idea of neural networks. In order to access their scope, we want to introduce three minimum requirements for a meaningful QNN that is based on the Hopfield Neural Network model and contains the feature of associative memory. These requirements can be summarised as: 
\begin{enumerate}
\item The initial state of the quantum system encodes any binary string of length $N$. The QNN produces a stable output configuration encoding the one state of $2^M$ possible output binary strings of length $M$ which is closest to the input by some distance measure.
\item The QNN reflects one or more basic neural computing mechanisms (attractor dynamics, synaptic connections, integrate \& fire, training rules, structure of a NN)
\item The evolution is based on quantum effects, such as superposition, entanglement and interference, and it is fully consistent with quantum theory.
\end{enumerate} 
The first point ensures that the QNN has the feature of associative memory, pattern recognition and other central properties of neural information processing. Requirement $(2)$ demands a connection to the idea of neural computing, but is held very general in order to cater for the variety of existing and future approaches. It is still important in order to exclude quantum computing algorithms that simulate associative memory with no relation to neural networks, so called \textit{Quantum Associative Memories}. The third point makes sure that the network qualifies as a \textit{Quantum} Neural Network. Of course, there are other possible frameworks for QNNs. For example, there is theoretically no reason to confine the information processed to binary units as given in classical computers. However, these requirements shall serve as a guideline to access existing proposals of QNNs. Also note that the requirements include Feed-Forward NNs and other typical tasks such as pattern classification and pattern completion in addition to HNNs and associative memory which are analysed here.\\

If confronted with the problem of finding a QNN model that fulfils points 1-3, a fundamental challenge appears in the integration of the nonlinear, dissipative dynamics of attractor-based NNs and linear, unitary quantum theory. The problem becomes most apparent if we have a look at the perceptron setup. In neural computing, the incoming signal $\sum_{j=1}^N w_{ji} x_j $ of the $N$ neurons $x_j, \; j \neq i$ connected to neuron $x_i$ by weights $w_{ij}$ gets mapped to the output of neuron $x_i$ by a step-function, or, in case of graded-response neurons, by a sigmoid function. This nonlinearity leads to the attractor-like dynamics fundamental for associative memory. If we introduce quantum mechanics, we necessarily look at firing probabilities instead of deterministic values. However, mapping firing probabilities onto one another by a nonlinear function contradicts the basic principle of the linear evolution in quantum theory, where linear superpositions of solutions of the Schr\"odinger equation play an important role. One apparent exception are measurements, which can be understood as a probabilistic step-function process: a measurement would `collapse'  the superposition of a quron state onto one of the basis vectors $\{\ket{0}, \ket{1}\}$ with probability $|\alpha|^2, |\beta|^2$ respectively. However,  if a quantum equivalent of a perceptron would be constructed to simply induce measurements, the dynamics would evolve classically (more precisely, as a probabilistic classical perceptron) as quantum mechanical effects would be destroyed in every updating step. A more advanced approach is therefore required. \\

The idea to understand the neural activation function in terms of a quantum measurement has been attempted by several authors, and seems an elegant solution to the problem to unify the diverse dynamics of neural networks and quantum theory. However, so far there has been no proposal that is able to capture the property of associative memory as demanded in Requirement $(1)$.  Other authors sacrifice the proximity to neural networks (demanded in $(2)$) or use quantum mechanics as an inspiration rather than implementing the full theory (violating Requirement $(3)$). The next section will review the different approaches to find a QNN model on the background of the three requirements in more detail.

\section{Review of existing approaches to QNNs}\label{rev}

\begin{figure}[t]
	\begin{minipage}[b]{0.49\textwidth} 
	\includegraphics[width=\textwidth]{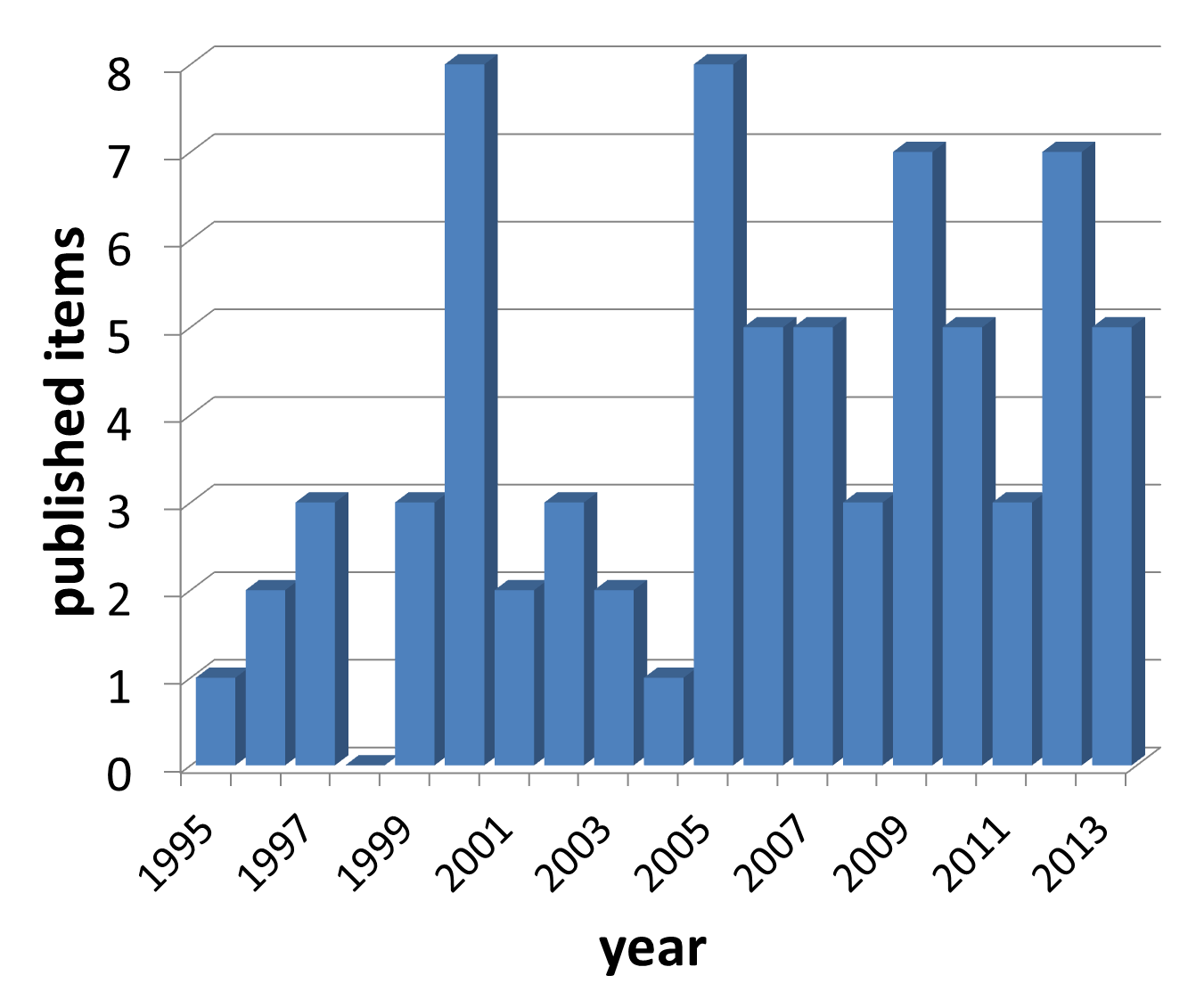}
	\end{minipage}
	\hfill
	\begin{minipage}[b]{0.49\textwidth}
	\includegraphics[width=\textwidth]{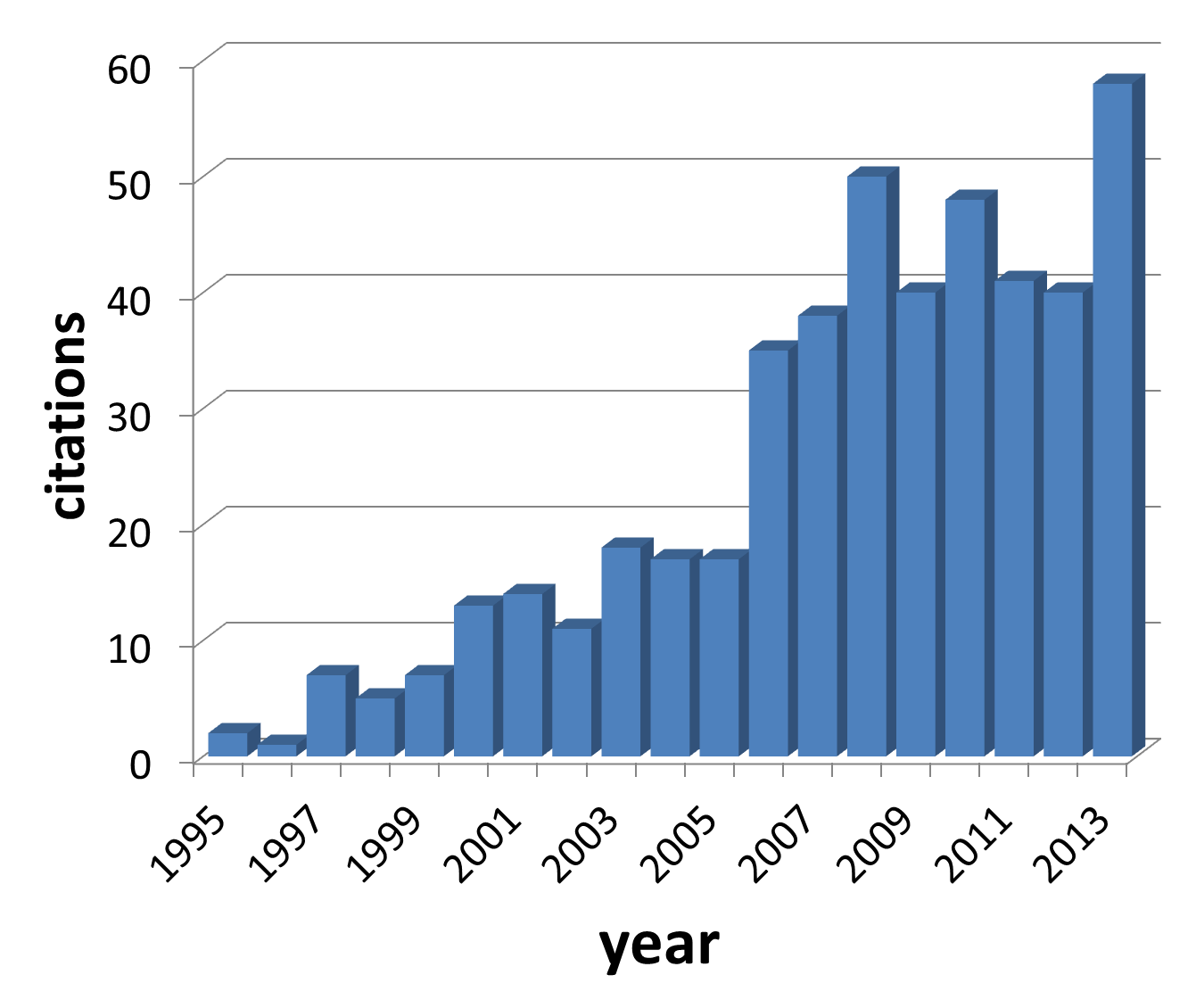}
	\end{minipage}
   \caption{Number of published articles (left) and citations of articles (right) with the term `Quantum Neural Networks' appearing in the topic or title since 1994 (Source: Thomsen Reuter's Web of Science).  Note that hits on neural network methods in quantum chemistry were manually excluded}
\label{stat}
\end{figure}

As mentioned earlier, the term `Quantum Neural Network' is claimed by a number of different proposals, and the field is still in the process to establish a coherent definition of its own subject. Fig. \ref{stat} shows the \textit{Thomsen Reuter's Web of Science} citation statistics for articles related to the keyword `Quantum Neural Network' that the number of publications has been steadily low, although the interest in QNNs is significantly growing. \\

The existing approaches can be summarised into a number of classes sharing a similar idea. The above discussed interpretation of QNNs on the basis of quantum measurements is an idea suggested in the early works on QNNs, and often remained a theoretical consideration rather than a fully functioning model \cite{kak95,menneer95,perus00,zak98}. A more practical approach that has received a relatively high amount of attention was Elizabeth Behrman's suggestion to realise a QNN through interacting quantum dots \cite{behrman99,toth00,faber02}. A large share of the literature on QNNs also comes from the discipline of quantum computing and tries to find specific quantum circuits that integrate the mechanisms of neural networks in some way \cite{gupta01,oliveira08,silva10,panella11,zhou12,goncalves14}. \textit{Quantum Associative Memories} \cite{ventura00,trugenberger01,andrecut03} are quantum algorithms that reproduce the properties of a neural network without trying to copy its dynamics or setup. Often cited is also the idea to introduce a \textit{Quantum Perceptron} \cite{siomau12,altaisky01}. In addition to these branches, a number of other proposals will be mentioned. 

\subsection{Ideas to interpret the step-function as measurement}

Probably the first quantum approach to neural network research was published by Subhash K. Kak \cite{kak95}. He introduced the idea of ``quantum neural computation'' by interpreting the necessary condition for a stable state $x^0 = (x_1^0,...,x_N^0)$ in a `Hopfield-like' network defined by the weight matrix $w$ with entries $w_{ij}$,
\[ \mathrm{sgm}(w x_0 ) = x_0, \]
as an eigenvalue equation of a quantum system, $\hat{w}\ket{x_0} = \lambda \ket{x_0}$. The weight matrix $w$  thereby becomes an operator with eigenvector $\ket{x_0}$ and eigenvalue $\lambda =1$. Updating a network then corresponds to a quantum measurement that selects the eigenstates of a system. Kak notes that the sigmoid function is a nonlinearity that has no equivalent in the quantum mechanical formalism but does not discuss this crucial point any further. He concludes that ``brains are to be viewed as quantum systems with their neural structures representing the classical measurement hardware'' \cite{kak95}.\\

Mitja Peru{\v{s}}  \cite{perus00} followed similar lines  when he emphasises the analogy between what he calls a `Hopfield-like' network and quantum theory, comparing an updating function stripped bare of the important threshold or sigmoid function,
\[ x_i(t_2) = \sum\limits_{i=1}^{n} w_{ij} x_j(t_1), \]
 with the time evolution of a quantum  state,
\[ \psi (r_2, t_2) = \int \int G(r_1, t_1, r_2, t_2) \psi(r_1, t_1) dr_1 dt_1 . \]
Here, $G$ is a projection operator  in position representation defined as $G = \sum_{l=1}^{k} \psi^{l*} (r_1, t_1) \psi^l(r_2, t_2)$, $r_1, r_2$ are position and $t_1, t_2$ time variables, and \{$\psi^l(r,t)$\} is a complete basis of the quantum system. $G$ shows an intriguing analogy to Hebb's learning rule Eq. (\ref{hebb}). Peru{\v{s}} also models a pattern recall through a collapse of the wave function. A valid input state $\psi$ is supposed to be almost orthogonal to all memorised vectors (or basis states) $\psi^l $ except from one vector $\psi^j$ ($l\neq j$), so that $G$ applied to the input retrieves $\psi^j$ with a high probability $|\psi\psi^{j}|^2$.  Peru{\v{s}}' approach has been further developed by ideas of interacting quantum dots presented below.\\

Shortly after Kak published his thoughts on quantum neural computation, Menneer and Narayanan introduced their version of a \textit{Quantum Inspired Neural Network} \cite{menneer95}. The basic idea, taken from the many-universe interpretation of quantum mechanics, is to look at a `superposition of networks' each storing one pattern (instead of one network storing several patterns). Such a `network quantum state is given by its weight vector. Retrieving a pattern corresponds to the collapse of the superposition of weight vectors and thus ???choosing??? one network which in turn retrieves the desired result. Menneer and Narayanan describe a collapse mechanism that results in choosing the one network whose stored pattern resembles the input pattern most. \\
\begin{figure}
\centering
                \includegraphics[width=0.6\textwidth]{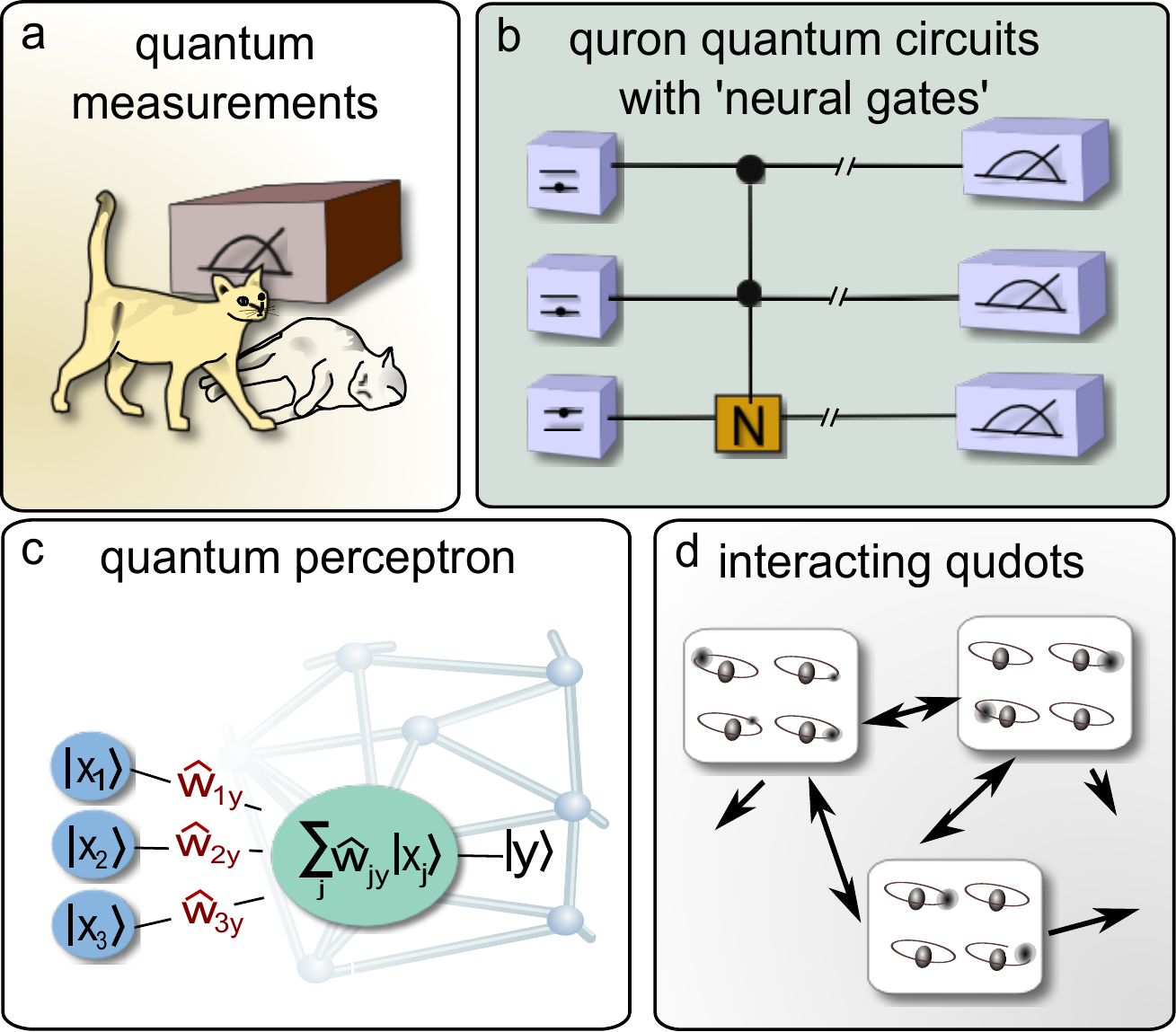}
            \caption{Different approaches to develop a Quantum Neural Network. (a) Several authors notice the analogy between the nonlinear stepfunction in the activation of a neuron and a measurement process, here symbolised by Schr\"odinger's famous cat. (b) Many contributions try to build quantum circuits with gates or features inspired by neural computing. (c) A challenge in finding a QNN model is to translate the core mechanism of a perceptron into a corresponding quantum version without loosing the rich dynamics of neural networks. (d) An interesting branch of proposals understands interacting quantum dots consisting of four atoms sharing two electrons as a QNN.  }
  \label{second}
\end{figure}

A largely unnoticed but nevertheless interesting QNN approach along the lines of quantum measurements has been proposed by Zak and Williams \cite{zak98}. The authors try to find a quantum formalism that captures the two main properties of Hopfield networks, dissipation and nonlinearity, by replacing the step- or sigmoid-activation-function with a quantum measurement.  Zak and Williams do not consider single neurons as quantum objects, but introduce a unitary walk between the quantum network basis states $\{\ket{0,...,0}_1,..., \ket{1,...,1}_{2^N}\}$. The evolution maps the amplitude vector  
\[a = (a_1, ..., a_{2^N} ),\]
where $a_i$ is the amplitude of the quantum network basis state $\ket{x_1,...,x_N}_i$ as given in Eq. (\ref{nws}), onto a vector $a'$ by applying a unitary transformation $U$:
\[a' = Ua \]
This is followed by a projective measurement $\sigma$, collapsing the superposition from Eq. (\ref{nws}) onto one of the network basis states 
\[(a_1, ..., a_{2^N} ) \; \; \rightarrow \;\; (0_1,...,1_i, ..., 0_{2^N}), \] 
with probability $|a_i|^2$. This map is ``nonlinear, dissipative, and irreversible, and it can play the role of a natural  `quantum' sigmoid function'' $\sigma$  \cite{zak98}. The full network dynamics in an update from timestep $t$ to $t+1$ are consequently given by
\[a'_{t+1} = \sigma(Ua_t).\]
Here, the evolution matrix $U$ is of dimension $2^N \times 2^N$ and contains the probabilities of transitions between network basis states $\ket{x_1,...,x_N}$. The operation is required to be unitary. The authors unfortunately do not give a derivation of $U$ from a given Hopfield network that shows its characteristic dynamics. A major problem in such approaches is also the global coherence between the qurons. Even though Zak and Williams remark that the coherence only has to be kept up between two measurement processes for a sufficiently small time window $\Delta t$ \cite{zak99}, the high dimension of neural networks simulating brain functions make it a challenge to maintain global coherence.\\

In summary, the idea to achieve the NN dynamics through quantum measurements has been followed from different perspectives. However, the proposals all remain in the early stage of a suggestion and do not entail a fully developed QNN model with the required dynamics. In this way they all fail to contain the Requirement $(1)$ from above, demanding the property of associative memory. However, the idea to use quantum measurements to simulate the nonlinear convergence of associative memories seems to be the most mature solution to the dynamics incompatibility  problem encountered in the quest for a QNN model. 

\subsection{Interacting quantum dots}

Peru{\v{s}}' notion of the formal similarity between the updating function and the evolution of a quantum state was turned into a full QNN proposal by Elizabeth Behrman and co-workers \cite{behrman00}. The Green function was reformulated by a Feynman path integral which sums over all possible paths $\phi(t)$ propagating the system from $\phi^0$ to $\phi^{'}$
\[\ket{\psi}_{T,\phi^{'}} = \int D\phi(t) \exp\{\frac{i}{\hbar} \int \frac{m}{2} \dot{\phi}^2 -V(\phi,t) dt\} \ket{\psi}_{0,\phi^0 }.\]
Here, $V$ describes the potential and $m$ the mass in the quantum system. Instead of different qurons, the network is realised by the propagation of one quron only. Its state after each of $N$ different time slices simulates the states of $N$ virtual neurons. The synaptic weights are then engineered by the interaction of the 2-level quron with an environment. In other words, instead of updating the state of a neuron $x_i$ ,$i \in \{1,...,N\}$ , by the state of $N-1$ others, the state of one neuron at time $i\Delta t$ is -- in different intensities-- influenced by its state at times $0,...,(i-1)\Delta t$ before. Behrman et al. propose to implement this `time-array neural network' through a quantum dot molecule interacting with phonons of a surrounding lattice as well as an external field that leads to a potential $V(r)$. The network can be trained by a common backpropagation rule, in which the `weights' are corrected according to the output error gradient. \\

Faber and Giraldi \cite{faber02} discuss Behrman's approach in the context of solving the incompatibility problem between neural and quantum computing. They ask the question of how the neural network's nonlinear dynamics can be simulated by a quantum system. As remarked by Behrman before, they work out that the nonlinearity arises naturally in the time evolution of a quantum system through the exponential function and the nonlinear kinetic energy that feeds into the potential $V(\phi, t)$. However, Behrman and co-workers mention that for larger computations than simple gates, including more advanced properties of Hopfield networks like pattern recognition (and consequently also associative memory), a spatial-array model is needed. Such a model, using $N$ quantum dots has priorly been investigated \cite{behrman99,toth00,faber02}. It shows remarkable features such as the computation of entanglement or its own phase \cite{behrman13,behrman02}, but has not yet been implemented as a quantum associative memory device. \\

Behrman et al.'s proposal shows that the natural evolution of a system of interacting quantum dots can serve as an quantum neural computer in that the desired mapping of an input to an output can be engineered under certain conditions. In this manner, any quantum system with an evolution dependent on the initial state and with appropriate parameters is some kind of analog computing device for a given problem, just as any input-dependent physical system can be seen as an analog computer. It is therefore disputable if Requirement $(2)$ is sufficiently met. Although being a natural candidate for a QNN, the interactions between quantum systems still act very different to neural perceptrons, and to find the specific parameters that would lead to dynamics corresponding to a fully functioning QNN is by no means trivial.  

\subsection{Quantum neural circuits}

Some authors view the problem to find a QNN from a quantum computing perspective. Their constructions of complex quantum circuits are inspired by the dynamics of neural networks. For example, quantum circuits for qubits in which each quantum computation operation executed by unitary operators $\hat{U}$ is followed by a dissipative operator $\hat{D}$ were proposed \cite{gupta01}. These dissipative gates map the quantum state amplitudes onto $c\in \mathbb{C}$ or $0$, depending on whether it exceeds a threshold $\delta$. The operators $\hat{D}$ consequently mimic the perceptron activation function. Unfortunately, the authors do not give an example of a quantum system that could serve as such a gate. Faber and Giraldi \cite{faber02} suggest Behrman's realisation of a QNN to serve as the nonlinear dissipative operator $\hat{D}$.   \\

More recent approaches show only few similarities to neural network dynamics. Panella and Martinelli \cite{panella11} suggest general nonlinear quantum operators to construct a feed-forward network in which qubits are successively entangled with one another. Oliveira and Silva, together with their coworkers, develop a quantum logical neural network model that is based on classical weightless neural networks in which neural updating functions are stored in a table similar to a Random Access Memory on a computer  \cite{oliveira08,silva10,silva12}. A binary input string would `address' one of the registers and lead to a certain output. This model contains the ability of NNs to be trained by a learning algorithm, however it does not obtain the nonlinear convergence dynamics of ordinary neural networks. 

\subsection{Quantum Associative Memory (QAM) models}
% Algo
Quantum Associative Memories are quantum computing algorithms that simulate the property of associative memory without intending to use features of neural networks. This means that upon initialisation with an input pattern, a QAM quantum circuit selects the `closest' memory pattern in terms of Hamming distance. Most contributions in this category are based on Ventura and Martinez' important proposal \cite{ventura00} (with modifications \cite{trugenberger01,andrecut03, zhou12}). The basic idea is to run a quantum algorithm on a a superposition of all memorised states $\ket{M}$ that upon final measurement retrieves the desired output state with a high probability. Let $X^{P} = \{ \ket{x_1^{(1)},...,x_N^{(1)}},...,\ket{x_1^{(P)},...,x_N^{(P)} }\}$ again be the $P$ patterns to be stored in the QAM'. The memory superposition reads
\begin{equation} \ket{M} = \frac{1}{\sqrt{P}} \sum\limits_{p=1}^{P} \ket{x_1^{(p)},...,x_N^{(p)}}. \label{M} \end{equation}
An algorithm to create $\ket{M}$  is given in \citep{trugenberger01, ventura00}, and a more efficient version has more recently been introduced in \citep{long01}. A promising alternative would be to use the Grover search algorithm to create the memory superposition \cite{boyer96}. Since the result is probabilistic and the state `destroyed' by measurement, it is necessary to construct a number of duplicates of  $\ket{M}$. \\

The retrieval of a memorised state initially proposed by Ventura and Martinez  is based on Grover's famous algorithm for the search of a pattern in a superposition of a full basis \cite{ventura00}. Grover's original algorithm has to be modified to work on the initial state $\ket{M}$ as it only contains a selection of basis states. The resulting algorithm is able to retrieve all patterns that contain a certain input sequence. Ventura and Martinez consequently deal with the related problem of pattern completion rather than associative memory. Trugenberger \cite{trugenberger01, trugenberger02} uses a promising approach to render Ventura and Martinez' pattern completion algorithm into associative memory. He finds an evolution that writes the Hamming distance between the input state and each memorised state into the phase of the memorised states. This is essentially realised by a time evolution of $\ket{M}$ with the Hamiltonian 
\[ \mathcal{H} = \sum\limits_{i=1}^{n} \frac{(\sigma_3)_{i} + 1}{2}.\]
Here $\sigma_3$ is the third Pauli-matrix measuring the state of neuron $x_i$, $i \in \{1,...,N\}$, in the network state. A final measurement will then retrieve the patterns with a probability depending on their Hamming distance to the input. Repeated measurements identify the `closest' pattern to the input state. \\

The main advantage of a Quantum Associative Memory compared to classical Hopfield associative memory is the fact that it is theoretically able to store $2^N$ patterns into a qubit system of dimension $2n+2$ (whereof $n+2$ qubits serve as auxiliary units). Compared to the approximately $0.138$ patterns storable in a Hopfield network of $N$ neurons, this is a major improvement. However, the requirements of the algorithms proposed are still far beyond the status quo of current realisations of quantum computing. In addition to that, QAMs are no QNN models in the strict sense demanded here, since they are not based on neural network models and thus fail to meet Requirement $(2)$.\\

\subsection{Quantum Perceptrons}
The core of HNNs lies in its basic units that define how the state of a neuron is calculated depending on inputs from other neurons. An important question when developing a QNN model is thus how a `Quantum Perceptron' can be formulated. Two proposals tackle this problem of describing perceptrons with a quantum formalism. The first was Altaisky's \cite{altaisky01} introduction of a Quantum Perceptron. The perceptron is modelled by the quantum updating function
\begin{equation}\ket{y(t)} = \hat{F} \sum\limits_{i=1}^{m} \hat{w}_{iy}(t) \ket{x_i},  \label{altaisky}\end{equation}
with $\hat{F}$ being an arbitrary quantum gate operator and $\hat{w}_{ij}$ operators representing the synaptic weights, both working on the $m$ input qubits . The perceptron can be trained by the quantum equivalent of a learning rule :
\[w_{jy}(t+1) = w_{jy}(t) + \eta(\ket{d} - \ket{y(t)}) \bra{x_i}.\]
Here $\ket{d}$ is the target state and $\ket{y(t)}$ the state of neuron $y$ at the discrete time step $t$, and $\eta \in [0,1]$ the learning rate. Altaisky notes that this learning rule is by no means unitary regarding the entries of the weight matrix $w$ (for a proof see \cite{silva12}). The update would then fail to preserve the total probability of the system. A unitary learning rule however would not mirror the dissipative nature of learning \cite{altaisky01}. The author leaves this conflict unresolved. Some authores take this model further and show how the Quantum Perceptron is able to compute quantum gate operations like the Hadamard \cite{fei03} or C-NOT \cite{sagheer13} transformation.\\

Siomau \cite{siomau12} recently introduced a very different model of a Quantum Perceptron and tried to illustrate its powerful advantage of computing nonseparable problems\footnote{A problem is linearly separable if the respective outputs in phase space can be devided by a hyperplane. Perceptrons can only compute linear separable problems.}. His version does not use a quantum neural update function, but projectors $P = \ket{\psi}\bra{\psi}$ with the property $\braket{\psi}{x_1 \dots x_n} = |d| $, $d$ being the target output (the module $|\cdot |$ is taken to avoid unphysical outputs). He claims that the XOR operation for example can be computed by a two-input-neuron perceptron using the operator $P = P_{-1} + P_{+1} $ with $P_{-1} = \ket{00}\bra{00} + \ket{11}\bra{11} $ and $P_{+1} = \ket{01}\bra{01} + \ket{10}\bra{10} $. $ P_{-1}$ and $ P_{+1}$ are orthogonal and complete, which ensures that all inputs are unambiguously classified.\\

These ideas of a Quantum Perceptron may be taken as an inspiration that the core of a QNN lies in the construction of the basic units, which are Quantum Perceptrons. However, Altaisky's perceptron suffers from the lack of a definition of the Hilbert space that describes his quantum states $\ket{x_i}$. If (as defined here) the qurons each come from different Hilbert spaces, the operation of Eq. (\ref{altaisky}) would be ill-defined, since a quron is set equal to a sum of elements from other Hilbert spaces. Other definitions, like the direct sum of neuron Hilbert spaces, however would not entail the quantum property of superposition, and it would be difficult to maintain the claim of a `Quantum Perceptron'. Siomau's version also shows difficulties, since the operators $P$ fail to truly classify input states in a sense of writing information into the output state.

\subsection{Other approaches to QNNs}

Apart from these main branches of QNN research, there are a number of other ideas worth mentioning. Weigang \cite{weigang00} (and later Segher and Metwally \cite{segher09})  proposed a so called `Entangled Neural Network'. His setup is a set of subunits that are quantum teleportation devices \cite{nielsen10} between two `neurons' that are able to store, evolve and measure quantum information. The output of a teleportation process of a subunit is fed into the teleportation process of the next subunit. Weigang gives an example of encoding certain properties into the qubit's states, amplitudes and phases so that in the end an optimisation function can be retrieved.  Neigovzen et al. \cite{neigovzen09} use adiabatic quantum computing to successively change the Hamiltonian of a quantum system of a quantum object in a well potential into a NN-like Hamiltonian in which the memory states are represented by several energy minima, such that the object ends up in the closest minima to its initial state. A number of publications investigate fuzzy logic neural networks \cite{li02,purushothaman97,rigatos07}. Worth mentioning is an example of a QNN interpreting the synaptic weights as fuzzy variables updated by a fuzzy learning algorithm and thereby create an associative memory very close to quantum mechanical concepts \cite{rigatos07}. Fuzzy QNNs received a fair amount of attention compared to other QNN approaches, however, despite using the name `Quantum Neural Network' they do not respect quantum theory, but are inspired by a quron's continous range of coefficients in the interval $[0,1]$.\\

Other efforts have tried to make use of the Ising-type energy function describing the network's dynamics introduced above. Spin-glass models have been extensively used in the 1980s to analyse the thermodynamic properties of Hopfield neural networks \cite{amit85,hemmen86}. A widely recognised neurophysiological experiment on the salamander cortex gives evidence to believe that 'real' neurons are indeed connected by pairwise correlation as in an Ising-type model \cite{schneidman06}. Nishimori and Nonomura \cite{nishimori96} for example analyse a $x-y$ quantum Ising model simulating a neural network. They come to the conclusion that quantum fluctuations play the same role as thermal fluctuations and are consequently not able to explain or influence the macroscopic firing dynamics other than a classical model with nonzero temperature would do.

\section{Discussion}\label{dis}

This article reviewed the different approaches to find a Quantum Neural Network model. It introduced into neural computing as well as quantum computing. The incompatibility between the nonlinear, dissipative dynamics of the former and the linear, unitary dynamics of coherent quantum computing was pointed out. A framework for potential QNNs was established by introducing three requirements regarding $(1)$ the input-output relation of QNNs, $(2)$ the foundation in neural network theory and $(3)$ the use of and consistency with quantum theory. Existing proposals for QNNs were presented and evaluated against this framework. \\

As a conclusion, QNN research has not found a coherent approach yet and none of the competing ideas can fully claim to be a QNN model according to the requirements set here. The problem seems to lie in the above mentioned incompatibility between both dynamics. Either the dissipative dynamics of NNs is exploited to obtain the attractor-based feature of associative memory, which leads to a mere superficial application of quantum theory as in fuzzy logic neural networks. Or, as in most proposals, a quantum mechanical evolution is merely inspired by elements of neural network theory, for example given in the ideas of interacting quantum dots, neural quantum circuits or Quantum Associative Memory. The Quantum Perceptrons and quantum measurement proposals seem to give a more comprehensive solution to the problem, however, both fail to lead to the construction of a mature QNN model.  \\

It seems to be vital to attempt the quest for a QNN from the stance of a more advanced formulation of quantum theory. A candidate for a quantum system simulating a classical neural network's attractor-like dynamics would need to contain at least two stable states that are obtained through dynamics highly dependend on the initial conditions. Stable or equilibrium states are typical for dissipative quantum systems interacting with an environment, so called open quantum systems \cite{breuer02}. Besides the potential future developments of dissipative quantum computing \cite{verstraete09} mentioned above, an interesting perspective is given by the concept of Open Quantum Walks \cite{attal12,sinayskiy12,bauer13}. Open Quantum Walks are a tool to describe evolutions of a system's `external' quantum state due to an internal degree of freedom interacting with an environment. Their advantage compared to coherent Quantum Walks \cite{kempe03} that have been recently applied to QNNs \cite{schuld14} is not only the inclusion of an environment, but also the fact that no global coherence between the qurons is needed in order to exploit quantum effects \cite{zak98}. The output state, encoded in the internal degree of freedom of the walker, can be furthermore read out without destruction of the coherence by measuring the external degree of freedom \cite{attal12}. However, Open Quantum Walks are a fairly new tool and their dynamic properties and possible application to QNNs are yet to be studied.\\

Dissipation described by open quantum systems is not only an important factor to include to obtain the mathematical structure encountered in neural computing. Although QNNs do not directly claim to be `quantum brain models', they can be understood as part of recent developments to find quantum effects that optimise biological processes \cite{ball11}. Quantum biology, the new emerging field summarising these efforts, deals with quantum systems strongly coupled to a biological environment that induces decoherence and dissipation. If QNNs are supposed to be regarded as investigations into potential quantum effects in biological neural networks beyond computing devices, dissipative models including an environment will be necessary to consider. \\

\begin{acknowledgements}
This work is based upon research supported by the South African Research Chair Initiative of the Department of Science and Technology and National Research Foundation.
\end{acknowledgements}

%\bibliographystyle{unsrt}       
%\bibliography{bibliography}   

\end{document}